\documentclass[12pt]{article}
\usepackage{amsmath}
\usepackage{amsthm}
\usepackage{amsfonts}
\usepackage{amssymb}
\usepackage[dvips]{graphicx}

\input epsf

\newtheorem{thm}{Theorem}[section]

\newtheorem{com}[thm]{Comments}

\newtheorem{lemma}[thm]{Lemma}

\newcommand{\C}{{\mathbb C}}

\newcommand{\Z}{{\mathcal Z}}

\def\p{\partial}

\def\e{\epsilon}
\def\f{\frac}

\def\la{\longrightarrow}
\def\ol{\overline}

\def\us{\underset}
\def\s{\sigma}
\def\S{\Sigma}
\def\t{\text}
\def\tn{\textnormal}

\title{Quantum Computation and the localization of Modular Functors
\footnote{Based on lectures prepared for the joint Microsoft/University of
Washington celebration of mathematics April 2000 and the AMS meeting on mathematics in the new millennium UCLA, August 2000.}}
\author{Michael H. Freedman \footnote{Microsoft Research, One Microsoft Way,
Redmond, WA 98052-6399}}

\begin{document}

\maketitle

{\it  \small\ Dedicated to my teachers and collaborators: Alexei
Kitaev, Greg Kuperberg, Kevin Walker, and Zhenghan Wang. Their
work has been the inspiration for this lecture.}

\begin{abstract}
The mathematical problem of localizing modular functors to
neighborhoods of points is shown to be closely related to the
physical problem of engineering a local Hamiltonian  for a
computationally universal quantum medium.  For genus $=0$
surfaces, such a local Hamiltonian is mathematically defined.
Braiding defects of this medium implements a representation
associated to the Jones polynomial and this representation is
known to be universal for quantum computation.
\end{abstract}

\section{The Picture Principle}
Reality has the habit of intruding on the prodigies of purest
thought and encumbering them with unpleasant embellishments.  So
it is astonishing when the chthonian hammer of the engineer
resonates precisely to the gossamer fluttering of theory. Such a
moment may soon be at hand in the practice and theory of quantum
computation. The most compelling theoretical question,
$\lq\lq$localization," is yielding an answer which points the way
to a solution of Quantum Computing's (QC) most daunting
engineering problem: reaching the accuracy threshold for fault
tolerant computation.

After Shor's discovery [S1] of a polynomial time factoring
algorithm in the quantum model QC, skeptics properly questioned
whether a unitary evolution could ever be induced to process
information fault tolerantly. The most obvious tricks, such as
making a backup copy, useful in a dissipative system (e.g. pencil
and paper) are unavailable in quantum mechanics.  To overcome
these difficulties, a remarkable theoretical framework based on
$\lq\lq$stabilizer codes,"  $\lq\lq$transversal gates,"
$\lq\lq$cat-state-ancilli, " and nested concatenations of these
was erected [S2], [S3], [A,B-O], [K1], and [KLZ]. While the result
is a consistent recipe for fault-tolerant quantum computation, the
accuracy threshold which would allow this combinatorial behemoth
to overcome its own overhead has been estimated as about
$10^{-6}$, one i.i.d. error per one million physical gate
operations and requiring gates accurate also to one part in a
million.  This places a formidable task before the engineer and
physicist.  But within the year the beginnings of a new idea on
fault tolerance had been generated by Kitaev [K2].

While the term is not yet present in that paper the idea is to
construct (first mathematically) a $\lq\lq$quantum medium" and to
store quantum states as topological structures within the medium
and (eventually) manipulate these states, that is, apply gates to
them, by topological transformations of the medium.  For our
purposes, we define a quantum medium as a collection of many
finite level systems coupled together by a Hamiltonian $H$ obeying
a strong locality condition:  The individual systems are located
in a $2-$dimensional lattice or a more irregular cellulation of a
surface $\S$.  We postulate a constant $d > 0$ so that $H=\S
\ol{H}_k$ and each $\ol{H}_k= H_{k}\otimes$id,  where the
identity is on all tensor factors(= subsystem) not located within
some ball $B_{\ell}$ of diameter $d$ in the lattice.  For
example, the Heisenberg magnet with $H= -J\us{a,b\, =\, \partial\,
\t{edge}}{\S} \overset{\rightharpoonup}{\s}_a
\otimes\overset{\rightharpoonup}{\s}_b$ is a quantum medium of
diameter $=1$.  (But engineer be warned; localizing $H^\ell$
within balls of diameter $=d$ implies $n-$ary interaction for $n
\sim d^2$. Controlling effective $n-$ary terms for $n\geq2$ will
be tricky in the extreme and probably will require enforcing
symmetries to cancel lower order terms.)  Kitaev's $\lq\lq$toric
code" [K2] in which quantum states are stored as first homology
of a torus, can be counted as having $d=2$; they require $4-$ary
interactions.

We study here a partial generalization of the toric code which
also stores quantum information in a degenerate ground state
$V(\S)$ of a quantum medium.  The medium is on a disk with
point-like defects which we treat as punctures. The dimension of
$V(\S)$, $\S$ the punctured disk, grows exponentially with the
number of punctures.  Transformations of $\S$, that is
braidings (up to isotopy) of the punctures in space-time, $\S
\times R$, operate unitarily on $V(\S)$.  Other work
([K2], [P], and [K,B]) also explores the realization of elements
of computation by braiding anyonic $\lq\lq$quasi-particles" or
$\lq\lq$defects" of a quantum medium.

The vision is that stability of computation, at least sufficient
to reach the $10^{-6}$ threshold for $\lq\lq$software" error
correction, is to be realized by the discreteness of algebraic
topology:  two $Z_{2}-$homology cycles are never $\lq\lq$close,"
two words in the braid group are equal or distinct.  More
exactly, it is geometry not topology which will confer stability.
Working in a lattice model one may calculate [K2] that the
perturbation Hamiltonian $P$ must be raised to the length scale
$L$ before nonzero terms, $< \zeta |P^{L}|\eta>, \zeta, \eta \in$
ground state $(H)$, are encountered and so the splitting of the
ground state is estimated to be proportional to $e^{- \Omega(L)}$.
The length scale in the previous two examples are: $L=$ (length of
shortest essential cycle); and in the anyonic context, the
closest that two defects are allowed to come to each other during
braiding. The $\lq\lq$engineering goal" is to construct a
\underline{physical} quantum medium on a material disk whose
ground state admits many localized excitations ($\lq\lq$anyons")
whose braidings effect computationally universal unitary
transformations of the ground state.  It is further hoped that actual
$\lq\lq$errors," the result of unwanted noisy excitations, are to
be removed automatically by some relaxation process in which the
system is coupled to a cold bath by another much weaker
Hamiltonian $H^\prime$. The mathematicians first cut at the
engineering goal is to produce a \underline{mathematical} quantum
medium with these properties and this is accomplished by the
theorem below. This $\lq\lq$first cut" is not yet interesting to
experimentalists since the Hamiltonian contains summands which
have as many as $30$ nontrivial indices, but it represents an
exact existence theorem.  The question for physicist is whether
this phase can also be represented perturbatively with a simple
Hamiltonian, perhaps a RVB model [A], [N,S].  This would be a
major step toward physical realization.

\begin{thm}Consider a rectangle $R$ of Euclidian square lattice
consisting of $15$ boxes by $30\, n$ boxes.  Associate a $2-$level
spin system $\C^2$ with each of the $e :=960n + 36$ box edges in
$R$. The disjoint union of these spin systems has Hilbert space
$(\C^{2})^{\otimes e}=:X$.  There is a time dependent local
Hamiltonian $H_{t}=\left(\us{k}{\S} \ol{H}_{k, t}\right)$ with
fewer than $2000\,n$ terms and each $H_{k}$ having $30$ or fewer
indices, supported in at most a $5 \times 3$ rectangle of boxes -
$\lq\lq$diameter $=5$."  For $t=0$, the ground states of $H_0$
form a sub-Hilbert space $W \subset X$, and geometrically
determines $3n$ exceptional points or $\lq\lq$defects" spaced out
along the midline of $R$. Within $W$ there is a
$\lq\lq$computational" sub-Hilbert space $V\cong (\C^{2})^{\otimes
n}$,  $V\subset W.$ $\,W$ may be identified with the
$SU(2)-$Witten-Chern-Simons modular functor at level $\l=r-2=3$
of the $3n-$punctured disk with the fundamental representation of
$SU(2)$ labeling each of the $3n+1$ boundary components.  The
Braid group $B(3n)$ of the defects acts unitarily on $W$ according
to the Jones' representation at level $=5$.  Any quantum algorithm
can be efficiently simulated on $V$ by restricting the action of
$B(3n)$ to a $\lq\lq$computational subspace."

The representation is implemented adiabatically by gradually
deforming $H_t$ to $H_{t+1}$ and then to $H_{t+2}$ and so on. The
passage from $H_t$ to $H_{t+1}$ involves turning off an
exceptional term $\ol{H}_{k, t}$ which defines a defect site and
turning on a new term $\ol{H}_{k, tH}$, which determines an
alternative, adjacent, site for the defect at time $t+1$. Each
braid generator can be implemented in $4(r +1)$ times steps. We
believe, based on a conjectural energy gap, that the geometry
confers stability to this implementation which increases
exponentially, error = $e^{- \Omega (L)}$, under refinement of
the lattice on $R$ by a factor of $L$, while the number of time
step needed for a computation increases only linearly in $L$.

\end{thm}

\begin{com}\tn{\begin{itemize}
\item The second paragraph of the theorem should be
read as a defensible physical proposition, whereas the first
paragraph is mathematics.
\item Our Hamiltonian may be too
complicated to prove the persistence of an energy gap above the
ground state in the thermodynamic limit.  But based on an analogy
with a simpler system the gap is conjectured and will be
discussed at the end of the proof.
\item The passage from
the Jones' representation to computation on $V$ is the subject of
[FLW1] and [FLW2] where it is proved that universality holds for $r=5$ and
$r\geq7$. Functorially $V$ is a tensor summand of a subspace of
$W$ but by fixing a reference vector in the complementary tensor
factor we regard $V$ simply as a subspace of $W$.
\item The idea of anyonic computation is taken from [K2] and in a more speculative form [Fr].  The
new ingredient is the implementation of a computationally
complete modular functor by a local Hamiltonian. Witten's
approach [Wi] to CSr was Lagrangian and so nonlocal; it yields an
identically zero Hamiltonian under Legendre transform, [FKW] and
[A].  This lecture, in contrast, supplies a Hamiltonian
interpretation for CS5 (We may replace $5$ by any $r\geq7$ in the
statement at the expense of scaling the constants in the theorem
by $\f{r}{5}$ or $\f{r^{2}}{25}$ according to whether they scale
as lengths or areas.).
\item We know of two works in
progress with a similar objective.  Kitaev and Bravyi [K,B] study
a local model for the weaker functor CS4 on high genus surfaces,
and Kitaev and Kupperberg [K,K] have an approach to construct
local Hamiltonians  generally for modular functors on surfaces of
any genus which (unlike CS5) are quantum doubles [D]. Their
approach has the advantage that the local contributions to the
Hamiltonian can be arranged to commute so that an energy gap will
be rigorously established. In contrast, an interesting feature of the present
paper is that topological a combinatorial means yield an exact determination of a ground
state defined by \underline{non-commuting} terms.  This is not usually possible.
Finally, we will see that our local
construction for $H$ extends to the higher genus surfaces if CSr
is replaced by any modular functors of the form $V \otimes
V^{\ast}$.  The simple topological reason for this may illuminate
the analysis of [K,K].
\item Shortly, we will give the reader a completely pictorial understanding
of CSr on planar surfaces.
\end{itemize}}
\end{com}

So far, we have only discussed the $\lq\lq$engineering": the quest
to specify $H$ (which will be described in the proof). Let us
take a brief digression from that sulfurous underworld of
grinding gears to the Elysian fields of abstract thought.  The
Witten-Chern-Simons theory descends from the signature (=
Pontryagin form) in dimension 4 and every step of the desent to
lower dimension leads to deeper abstraction until mathematical
wit is well nigh exhausted as the point (dimension $=0$) is
reached.  To tell this story in its barest outline, we restrict
to $G=SU(2)$, and borrow from Atiyah [A], Freed [F], and Walker
[W].  The signature of a closed $4-$manifold is an integer as is
the Pontryagin class of an $SU(2)$ bundle over a closed
$4-$manifold.  An $SU(2)-$bundle over a closed $3-$manifold is
topologically trivial but if endowed with a connection acquires a
secondary $\lq\lq$ Chern-Simons" class in the circle $= R/ \Z$.
Quantizing [Wi] at level $\l$, leads to the topological
Jones-Witten-Chern-Simons invariant $\in \,\C$  which is morally
an average of the classical Chern-Simons invariant over all
connections.  The invariant for a closed surface $\S$ (with some
additional structure) is a finite dimensional vector space $V$;
and each $3-$manifold bounding $\S$ determines a vector $v\in
V$.  Before dividing by gauge symmetry, the vector space $\ol{V}$
is the infinite dimensional space of sections of the associated
complex line bundle to a natural $S^{1}-$bundle over the space of
$SU(2)$ connections $A$ on $SU(2)$ bundles over $\S$. A
$3-$manifold $Y$ with connection, $\ol{A}$, on a bundle extending
the bundle over the boundary, $\p (Y,\ol{A})=(\S,A)$, determines
a map $f\{(Y',\ol{A'})|\p(Y',\ol{A'})=(\S,A)\}\la S^1$ by
integrating the Chern-Simons form over $Y \cup -Y'$. The
consistent choices for such functionals constitute the total
space of this $\lq\lq$natural" $S^{1}-$bundle. In general, a map
$f$ is $\lq\lq$consistent" if it obeys the additivity properties
of the Chern-Simons integral: $f(Y')-f(Y'')=\tn{C.S.}(Y' \cup-
Y'')$. Symplectic reduction followed by quantization as explained
in [A] produces a finite dimensional $V$ from $\ol{V}$ with
$v(Y)\in V$ depending only on the topology of $Y$. The definition
of the Witten-Chern-Simons invariant for a surface with boundary
is a collection of vector spaces indexed by certain labelings.
For a $1-$manifold the invariant seems to be a certain type of
$\lq\lq2-$category" while the correct definition for a point is
but dimly perceived and the object of current research. Several
authors assert that it is unnecessary to finish the progression,
that we can be content with a theory whose smallest building
blocks are $\lq\lq$pairs of pants" (three-punctured- spheres).
The invariant for these while technically a vector in a
$2-$vector space is easily understood in terms of sets of vector
spaces parameterized by $\lq\lq$labelings" of the boundary
circles so no unusual categorical abstractions need be mastered.
The reason for this assertion is that using a handle body
decomposition all closed $3-$manifold invariants can be
calculated from gluing along surfaces with smooth boundary;
gluings along faces with corners on the boundary, which one would
encounter computing from a cellulation, can be avoided.  But the
Freed-Walker program rejects this advice on two grounds.  First
localizing $V(\S)$ not merely to $\lq\lq$pants," but to cells
(i.e. neighborhoods of points) may give more natural consistency
conditions, to replace the $14$ consistency equations of [W];
which in turn could eventually lead to classification of modular
functors and a conceptual understanding.  Second, to paraphrase
Edmund Hillary, we should localize to points $\lq\lq$because they
are there."

The hyperbole of the first paragraph can now be made sound.  CS5
is a universal model for quantum computation and for the
physicist/engineer to implement it, a local Hamiltonian $H$ must
be described.  For the pure mathematician to be satisfied with his
understanding of CS5 it must be localized to points. The two
objectives are certainly similar in spirit and possibly
identical.  To clarify the connection, we introduce an
intermediate concept, undoubtedly plebeian, but dear to a
topologist.  We would like when possible to describe a vector in
a modular functor as a linear combinations of $\lq\lq$admissible"
\underline{pictures} up to $\lq\lq$equivalence."  This, after all,
is exactly how we understand homology:  $v\in H_{1} (\S , Z_{2})$
is an equivalence class of admissible pictures.  To be admissible
the picture must be a closed $1-$manifold, the equivalence
relation is bordism.  Both $\lq\lq1-$manifold-ness" and
$\lq\lq$bordism" can be defined by local conditions which are the
combinatorial analogs of $\lq\lq$closed" and $\lq\lq$co-closed"
familiar from de Rham's theory of differential forms.  In Kitaev's
toric code these condition are imposed by vertex and face
operators $A_{v}$ and $B_f$ respectively.  There is a subtle
shift here from the usual way of thinking of homology as
equivalence classes of cycles to the $\lq\lq$harmonic"
representative which is merely the equally weighted average of
all cycles in the homology class.  In this way quotients and
equivalence classes are never encountered and homology is located
within cycles, within chains, just as a C.S.S. code space is
located within the fixed space of stabilizers built from products
of $\s_{z}$'s and further within the fixed space of stabilizers,
$\Pi \s_{x}$'s.

To generalize from homology, we should think of a picture as
(linear combinations of) anything we can draw on a surface $\S$.
If helpful, we allow various colors and/or notational labels,
framing fields, etc$\dots$, and even additional dimensions
bundled over $\S$. But in the present case no such embellishments
are required. What is important that if we move the surface by a
diffeomorphism, the picture should also move and move
canonically.  Thus if $\S$ is a torus it would not suite our
purposes to draw the picture of $v \in V(\S)$ in a solid torus
$T$, $\p T =\S$: a meridial Dehn twist on $\S$ extends over $T$,
twisting the picture, but a longitudinal Dehn twist does not have
any obvious way to act on a picture drawn in $T$. (To anticipate,
a modular functor will have an $S-$matrix which can transform a
picture in one (call it the $\lq\lq$inside") solid torus to a
picture in the dual ($\lq\lq$outside") solid torus where
longitudinal a Dehn twist does act.  But resorting to the
$S-$matrix does not solve our problem since its input and output
pictures are on a scale of the injectivity radius of the surfaces
and hence nonlocal.)  We demand that $\lq\lq$admissibility" and
$\lq\lq$equivalence" of pictures be locally determined, i.e.
decided on the basis of restriction to small patches on $\S$.  To
make the connection with lattice models, we consider $\S$
discretized as a cell complex; the conditions must span only
clumps of cells of constant combinatorial diameter.  As in the
example of harmonic $1-$cycles, $\lq\lq$equivalence" is a slight
misnomer: what we impose instead are invariance condition on the
(linear combinations of) admissible pictures representing any
fixed $v\in V$ which ensure that the stabilized vectors are in
fact equally weighted superpositions of all admissible pictures
representing $v$.

Now consider the question, perhaps the first question a geometric
topologist should ask about a modular functor $V (\S)$; Can you
draw a (local) picture of it on $\S$ so that the mapping class
group of $\S$ acts on $V(\S)$ by the obvious induced action on
pictures?

We should not expect it to be easy to discover the local rules for
the pictures associated to a given modular functor $V$ and in fact
they may not exist in much generality.  Recall that a three
manifold $Y$ bounding $\S$, $\p Y=\S$ determines a vector $v(Y)
\in V(\S)$ so we might think of our proposed picture $P\big(
v(Y)\big)$ drawn on $\S$ as some ghostly recollection of $Y$. The
present understanding of modular functors is closely related to
surgery formulas on links, but to think in this way we must
choose a $\lq\lq$base point" $3-$manifold $Y_0$ with $\p Y_0 = \S$
to hold the links.  This choice seems to create an asymmetry
which should not be present in $P\big(v(Y)\big)$. Thus for a
pictorial representation of $V$ which is derived from surgery, we
expect only part of the mapping group $-$ that part extending over
$Y_0 \,-$ will act locally.  To localize V, this problem must be
overcome.

Let us propose a meta theorem or $\lq\lq$principle" that solving
the $\lq\lq$picture problem," which we call $\lq\lq$combinatorial
localization," should imply both the Freed-Walker program, which
we call $\lq\lq$algebraic localization" and the design problem for
the Hamiltonian $H$ which we call $\lq\lq$physical localization."
\vskip.2in \epsfxsize=4.9in \centerline{\epsfbox{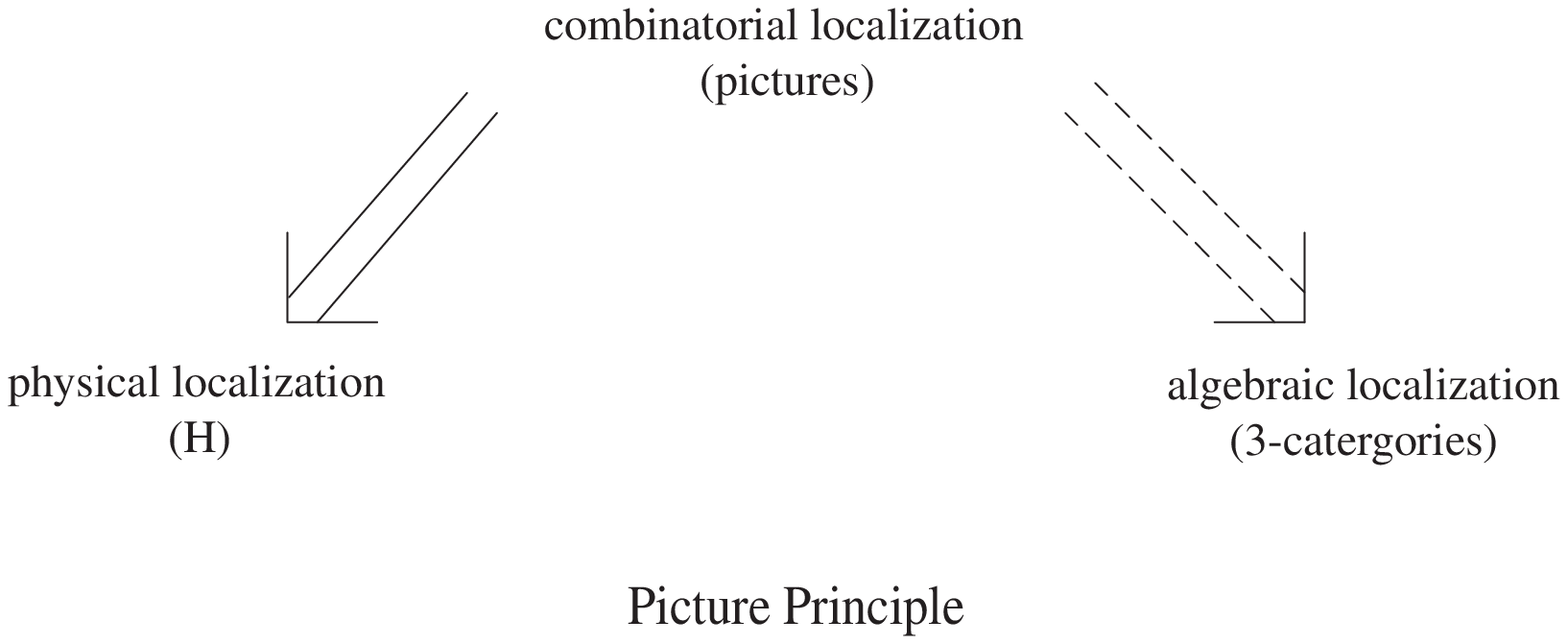}}
\centerline{Figure 1} \vskip.2in

The solid arrow is asserted with some confidence at least as a
mathematical statement; the dotted arrow is speculative.  While
the solid arrow seems unlikely to have a literal converse: ground
states of even simple Hamiltonians in dimension $\geq2$ are too
complicated to draw pictures of; conceivably the dotted arrow
might be an equivalence constituting a culmination of the
Freed-Walker program.

\section{Combinatorial localization of CS5 on marked disks, and
the proof of the theorem.}

We show how to represent CS5 (and by extension all CSr) on a disk
with marked points by local pictures.  Since the representation of
quantum computing within CS5 [FLW] only used the braid group
acting on a disk with marked points, this partial solution to the
combinatorial localization problem will suffice to prove the
theorem (once we have explained the solid arrow in figure 1).

For any $r\geq2$, CS$r$ has a combinatorial localization on any
cellulated  disk with marked labeled points, (labels $\e \{0,1,
\dots, r-2\}$ lie on the marked points and disk boundary) provided
the cellulation has bounded combinatorics and the marked points
stay sufficiently far from each other and the boundary.  For a
concrete statement, let us take the cellulated disk to be a
rectangle ${R}$ with a square Euclidean cellulation.  We
suppose that all marked points are at least $r$ lattice spacings
from the boundary and $9\,r$ from each other.  The marked points
and $\p R$ are all assigned the label $1$ (the irreducible $2$
dimensional representation of $sl(2, \C)_q$). In this circumstances
it is easy to build a trivalent $\lq\lq r-$collared rooted tree"
$T_r$ for the disk with marked points as shown in figure 2.

\vskip.2in \epsfxsize=3.5in \centerline{\epsfbox{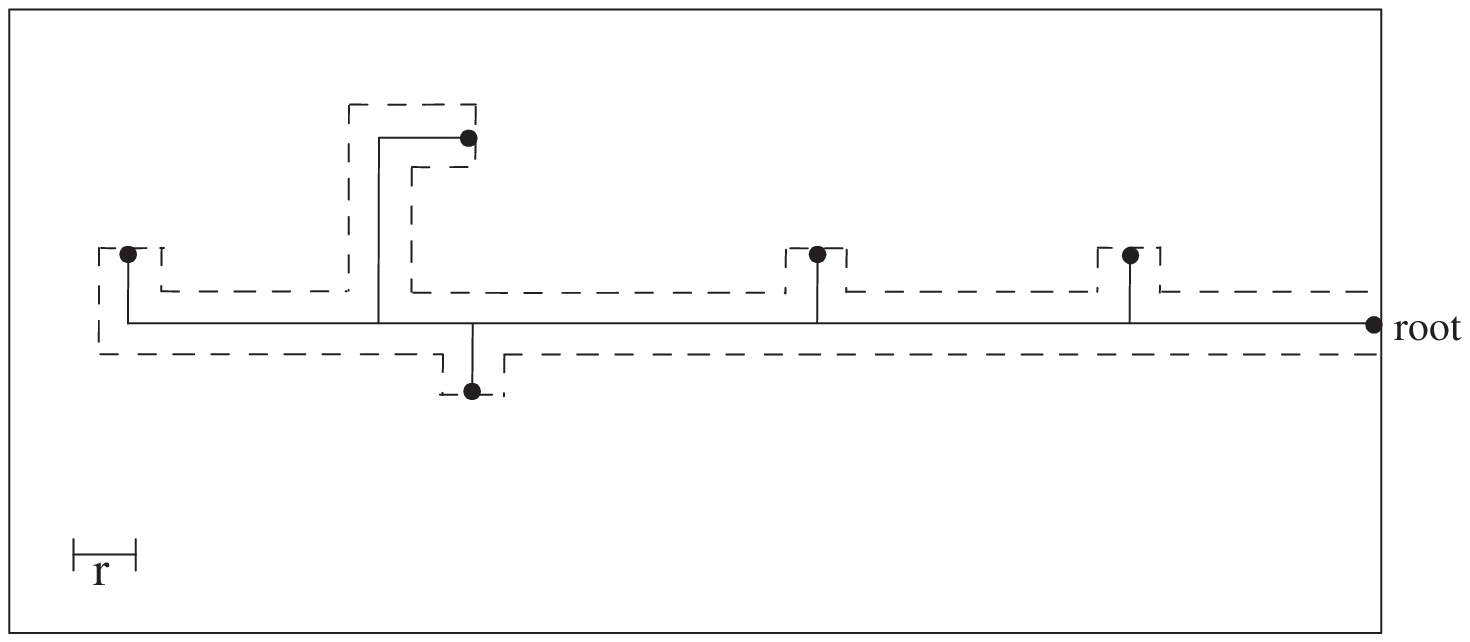}}
\centerline{Figure 2} \vskip.2in

All straight segments of the tree are to be more than $3r$
lattice bonds in length; the root is on $\p R$ and the leaves are
the marked points.  The $r-$collard condition is that an
$\lceil\f{r}{2}\rceil$ relative regular neighborhood $N(T)$ of
lattice cells - the region within the dashed line - should be
imbedded in $R$.

The box counts in the statement of the theorem are designed to
permit a (discontinuous) family of $T_r$'s to be found for at all
times during braiding. We say that the boxing of $R$ is
\underline{roomy} relative to the location of the marked points if it
has this property.  The key Lemma 2.1 will show that for roomy
boxing that two discrete pictures, which we regard as smoothly
equivalent are in fact combinatorially equivalent. More
precisely, the infinity of smooth averaging operators acting on
the space of combinatorial pictures has exactly the same joint
fixed set as a finite subset of combinatorial operators.

Let us begin with a geometric interpretation of CS5$(\S)=:V(\S)$.
For a closed surfaces $\S$ it is implicit in [K,L].  Let $\S$
bound a handle body $H$. A general $3-$manifold $Y$ with boundary
$\S$ can now be represented as a $\lq\lq$blackboard framed"
surgery diagram in $H$.  The special cabling morphism $w$ of the
Temperley-Lieb category (See chapter 12 [K,L] or [R,T]) when
composed into the surgery diagram yields a linear combinations
of $1-$ manifolds, each labeled by $\lq\lq 1$".  We may write $H$
as a planar surface cross interval, $H\cong\S\_\times I$, so that
$\S=\S\_ \, \cup_{\p}-\S\_$, where $-\S\_$ denotes $\S\_$ with its
orientation reversed. Now projecting these $1-$manifolds to
$\S\_$, we see a linear combination of immersed $1-$labeled
$1-$manifolds with overcrossings indicated at double points. This
pictures determines the vector $v(Y)$.  The Kauffman relations at
a root of unity, in our case $e^{2\pi i/5 }$, allow extensive
simplification of these pictures via the recoupling formalism. In
fact each $v \in V$ can be encoded as a labeling of a fixed
(framed, imbedded, and vertex planar) trivalent graph, which is a
spine for a $\S\_$.

It is an important observation of Walker's (personal
communication) and Gelca's [G] that this description can be
extended to labeled surfaces with boundary. (Verification follows
directly from the gluing axiom.)  In the case of a disk with $n$
marked points $(D, n)$ - treating marked points as crushed
boundary components - the modular functor with $n+1$ labels
$\overset{\rightharpoonup}{\ell} \t{,}\, \,
V_{\overset{\rightharpoonup}{\ell}} \,(D, n)$ has as its basis
$q-$admissible labelings with boundary condition on a fixed
trivalent tree imbedded in $D$, rooted on $\p D$, with leaves on
the marked points.  The boundary condition is that the label on
the root is the label given on $\p D^2$ and each leaf has the
label associated to its marked point. As in [FLW], we only need
consider the case where all labels $=1$.

The (framed) braid group acts on the labeled tree $T$ via its
imbedding in the disk.  To see the induced action on $V(D,n)$ (we
drop labeling subscripts), perturb the imbedding of $T$(rel its
endpoints) by pushing it downward into a three ball $D \times
[0,-1]$, where we think of $D$ identified with $D\times 0$. Now
implement any desired braid $b$ as a diffeormorphism of $D \times
[0,-\e]$ where $\e>0$ is small with respect to the previous push.
Viewed from above, $b(T)$ has overcrossings but the recoupling
$(6j)$ rules (and isotopies) allow $b(T)$ to be described in the
original basis of $q-$admissible labelings on $T$ (with root and
leaves still carrying the label 1). For example the simplest
Kauffman relations, on strands of $b(T)$ labeled by $\lq\lq1$"
read:
\[
\includegraphics[width=.45cm,height=.35cm]{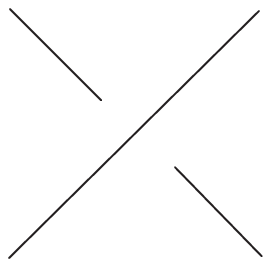}
= e^{\pi i/10} ) ( e^{-\pi i/10} \, \includegraphics[width=.55cm,height=.35cm]{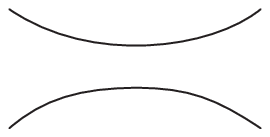}\tn{ and }
\, \bigcirc = e^{\pi i/5} +  e^{-\pi i/5} =:d.
\]
A detailed example: the effect of a single braid generator, is given immediately following the
statement of Lemma 2.1 to elucidate the recoupling of braids.

There is a topological observation inherent in inducing the braid
action on $V(D, n)$.  By capping off, any diffeomorphism of a
planar surface extends to the two sphere and can be extended
further to a diffeomorphism of the $3-$ball $B^3$.  The action on
$V$ comes from projecting this topological extension acting on
labeled trivalent trees back into the original planar surface
(after crushing the inner boundary components to points). In
fact, it is the correspondence between $3-$manifolds and diagrams
which proves that we have correctly specified the action on the
functor, for we have $v(\ol{f}Y)=f_{\ast}v(Y)$ where $\ol{f}|_{\p
Y=\S}=f$. Generally, when a surfaces $\S$ has genus $>0$ there
will be no way of including it in the boundary of a $3-$manifold
$M$ so that all diffeomorphisms of $\S$ extend over $Y$.  However
it is a triviality that any diffeomorphism of $\S$ extends over
$\S \times I$ by product with id$_{I}$.  Now let this extension
act on the appropriate equivalence classes of framed
$q-$admissibly labeled trivalent graphs imbedded in $\S \times I$
projected back into $\S$ to define the action  on any
$SU(N)-$level $=r$ modular functor $V$.  Thus the
$\lq\lq$doubled" functor $V(\S)\otimes V^{\ast}(\S)=V(\S\amalg
\ol{\S})=V \big(\p(\S \times I)\big)$ has a combinatorial
localization, i.e. is describable by local pictures.  This may
have some relation to unpublished work of Kitaev and Kupperberg
(private communication) on local descriptions for Drinfeld
doubles.

We set aside for later study the problem of devising combinatorial
local rules for the necessary elementary equivalences of such
trees $T$: $6j-$moves, ribbon equivalence, vertex half-twist
equivalence, and regular homotopy.

One would hope to define a quantum medium for CS5 of individual
systems with levels to record labels $0,1,2,3$ (and possible
additional levels to store other information) and terms $H_k$
with at most $6$ indices (as in a $6j-$symbol) corresponding to
these elementary equivalences. While this count seems correct in
the smooth setting, there the crude Hilbert space is infinite
dimensional which may create new difficulties.  We have not been
able to find a discrete setting in which all the equivalences are
expressed efficiently.  For the purpose of this lecture, we stay
with discrete models for quantum media built from $2-$level
systems, but to do this we accept terms $H_k$ with up to $30$
indices.

The fundamental $2-$dimensional representation of $SU(2)$
generates $SU(2)$'s complex representation ring and as a result
recoupling theory achieves a very simple result: an element $v \in
V(D, n)$ is a linear combination of imbedded $1-$manifolds each
labeled by $\lq\lq1$", i.e. the standard $2-$dimensional
representation and given the boundary condition: each
$1-$manifold of the linear combination meets each marked point
(and $\p D$) once.  Thus $\lq\lq$manifoldness" and the
$\lq\lq$boundary condition" define admissibility for our
picture.  this makes good sense combinatorically in the lattice
of $R$, as well as, smoothly.  We point out that our notion of
$1-$manifold is strict: at each vertex $0$ or $2$ edges (not $4$)
should be occupied.

It is time to define the local equivalence moves between pictures.
We are working within the Temperley-Lieb category modulo the
relation that the $(r-1)^{\t{th}} = 4^{\t{th}}$ Jones-Wenzl
projector is trivial.  This is our most interesting relation. As
a smooth equivalence relation this has only one form but
combinatorially, we need to impose two versions of it according
to how the output endpoints are grouped.  We denote these by
\includegraphics[width=.55cm,height=.50cm]{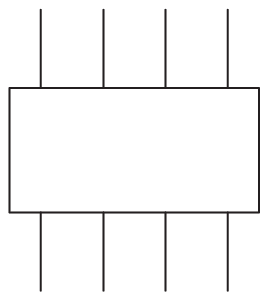} and
\includegraphics[width=.55cm,height=.50cm]{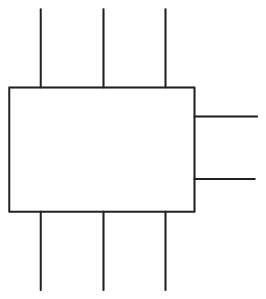}.  The second
picture stands for :
\includegraphics[width=.95cm,height=.85cm]{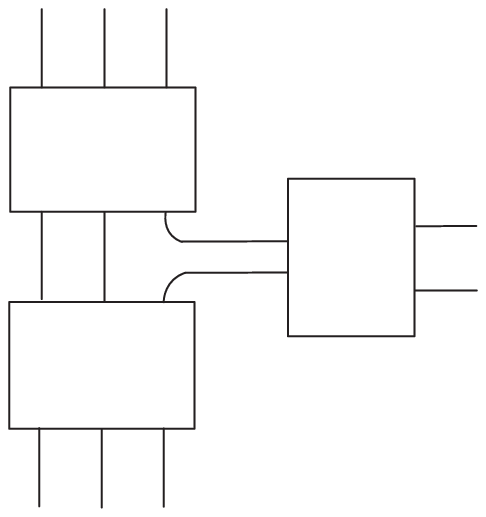} in conventional
projector notation ([K,L]).

A second relation says that removing
a circle which bounds a disk free from punctures multiples the
diagram by the scalar $\f{1}{d}$, $d =e^{ \pi i/5} + e^{-\pi
i/5}$. A third relation replaces the undercrossing that arise through
braiding with legitimate morphisms if the category.  In terms of
smooth pictures, the relation replaces the $\lq\lq$virtual"
uncrossing in the middle diagram with a two term sum:

\vskip.2in \epsfxsize=4in \centerline{\epsfbox{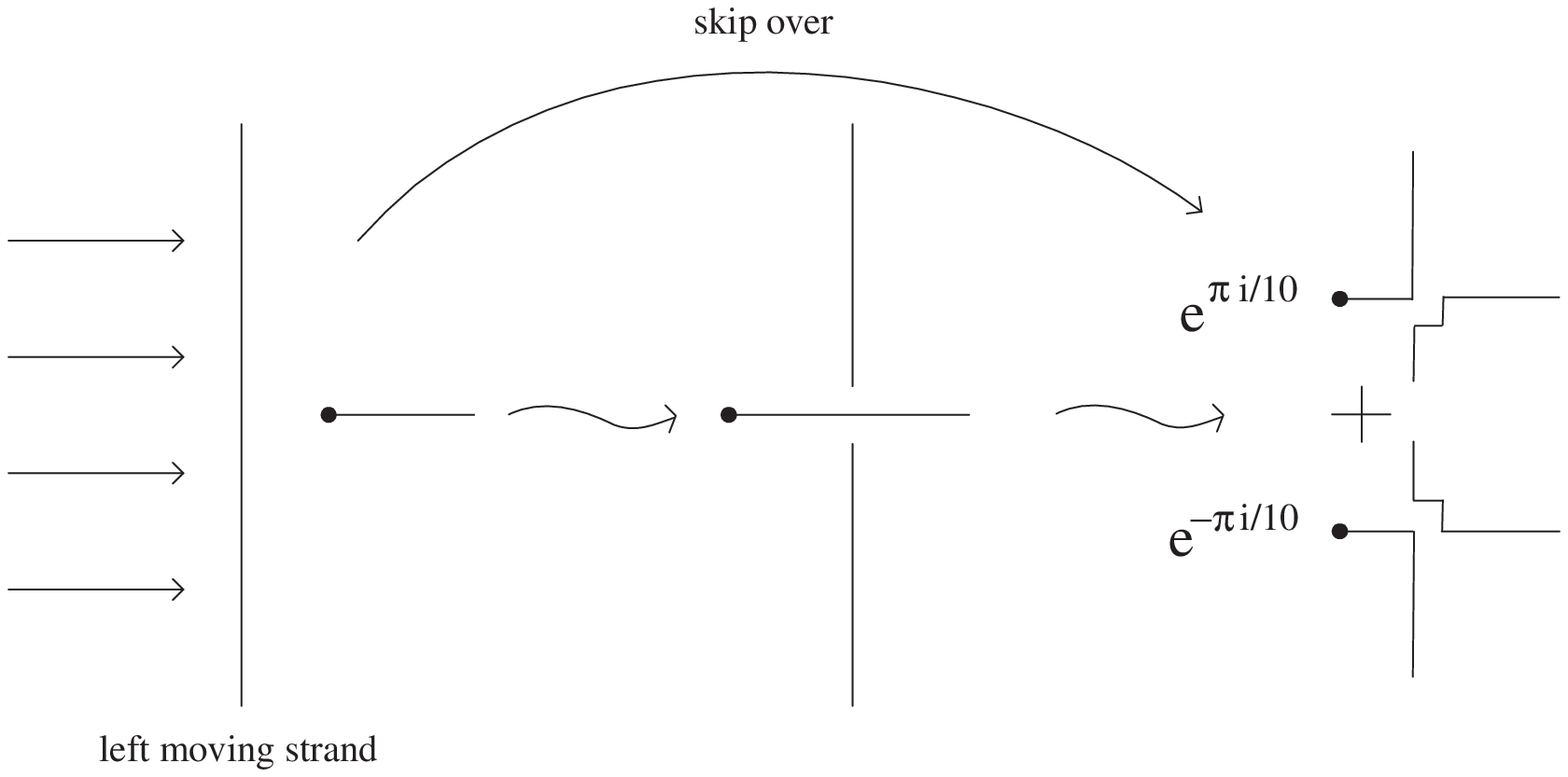}}
\centerline{Figure 3} \vskip.2in

The middle picture is $\lq\lq$ virtual"; it is not actually an
admissible picture to be assigned a weight. This relation requires
a little care and lattice space to discretize since we do not want
to permit the intermediate picture:

\vskip.2in \epsfxsize=1in \centerline{\epsfbox{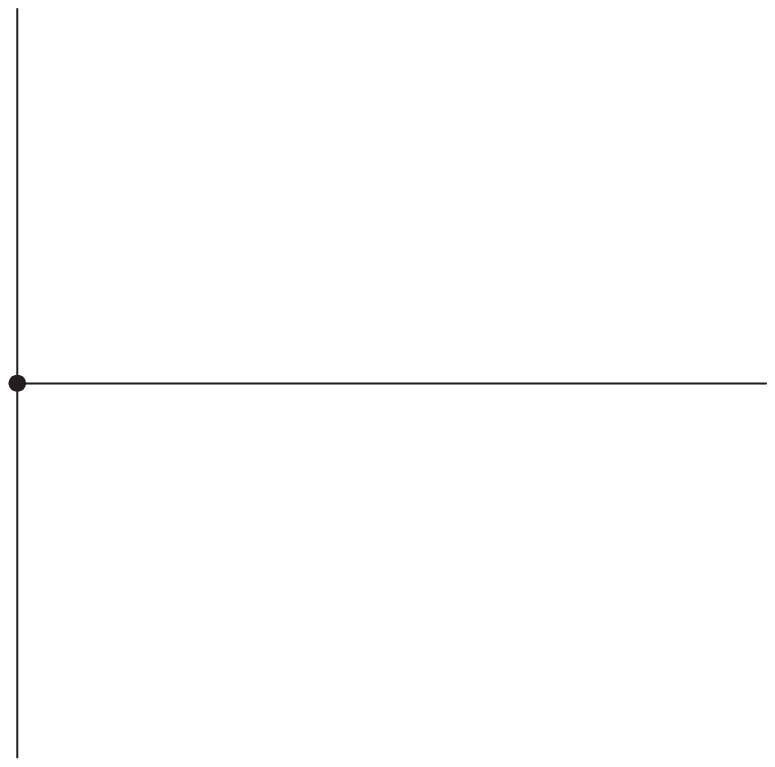}}
\centerline{Figure 4} \vskip.2in

\noindent which would represent the wrong boundary data at the
indicated defect.  Recall that each defect is labeled by $1$
representing the $2-$dimensional irreducible representation of
$sl(2, \C)_q$ which is recorded by a single line leaving the defect.

Finally, a fourth class of equivalence permits isotopy.  Again the
reader should note that enough neighboring sites should be observed by the
appropriate $H_k$ to preserve imbeddedness.  For example, cases 1
and 2 are allowable, case 3 is not.

\vskip.2in \epsfxsize=5in \centerline{\epsfbox{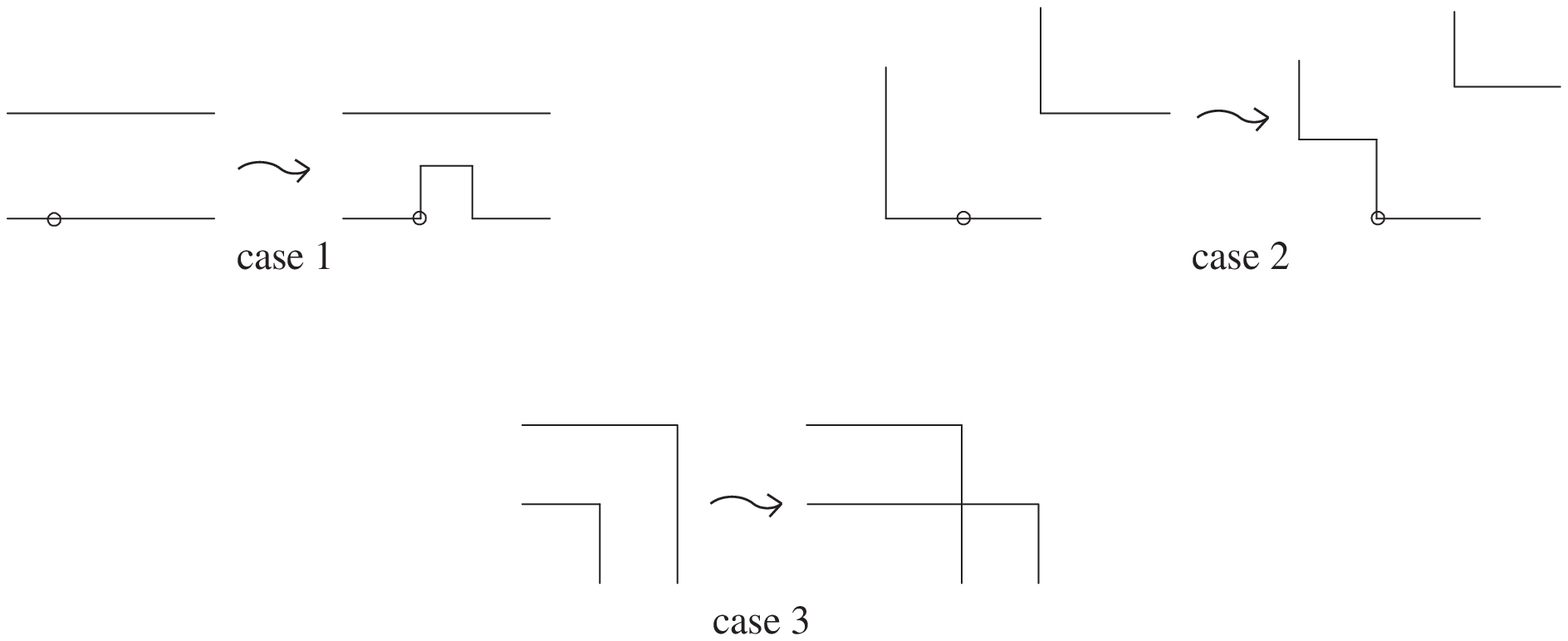}}
\centerline{Figure 5} \vskip.2in

There will be isotopy relations for arc endpoints as well.  For
example, in cases $1$ and $2$ of Figure 3 imagine the open circle
filled to become an end point and the shorter of the two line
segments meeting it deleted.  Morally, we should define operators
$\ol{H}_{k}$ which enforce the average of the initial $I$ and
final $F$ configuration of cases $1$ and $2$.  However there is a detail, to get the
overall phase correct, and not settle for merely a projective
representation, we must fix a base point direction: say the
positive ray emanating from each endpoint at $45$ degrees and
find positive semidefinite $\ol{H}_{k}$'s which assign zero norm
to $\f{1}{\sqrt{2}}(I_1 - e^{-\pi i/10} F_1 )$ and
$\f{1}{\sqrt{2}}[I_2 - F_2 ]$ in cases $1$ and $2$ respectively.
These operators correspond to asserting equivalences:  $I_1 \sim
- e^{-\pi i/10} F_1$ and $I_2 \sim F_2$.  The general rule is
that a state obtained by clockwise (counterclockwise) isotopy
through the base point direction must be adjusted by the phase $+
(-) e^{i \pi/2r}$ before being averaged.  Similarly there is an
isotopy relation for the arc end point on the boundary circle of
the disk $D$.  Here some point on the boundary is chosen an phase is
adjusted by $- (+)e^{i \pi/2r}$ as this point is crossed
clockwise (counterclockwise).

Let us return to the raltions \includegraphics[width=.60cm,
height=.50cm]{drawinga.eps} $=0 =$ \includegraphics[width=.60cm,
height=.50cm]{drawinga1.eps}.  Combinatorically the first may be
written out with the left hand side a $3 \times 3$ lattice square
foliated by parallel straight lines (of label $=1$). Wenzl's [We],
recursion formula, yields an identity equating $4$ parallel lines
with a linear combination of $13$ $\lq\lq$smaller" terms each
containing $\lq\lq$turn arounds."  The form  of the relation
\includegraphics[width=.55cm,height=.50cm]{drawinga.eps} is shown
below:

\vskip.2in \epsfxsize=4in \centerline{\epsfbox{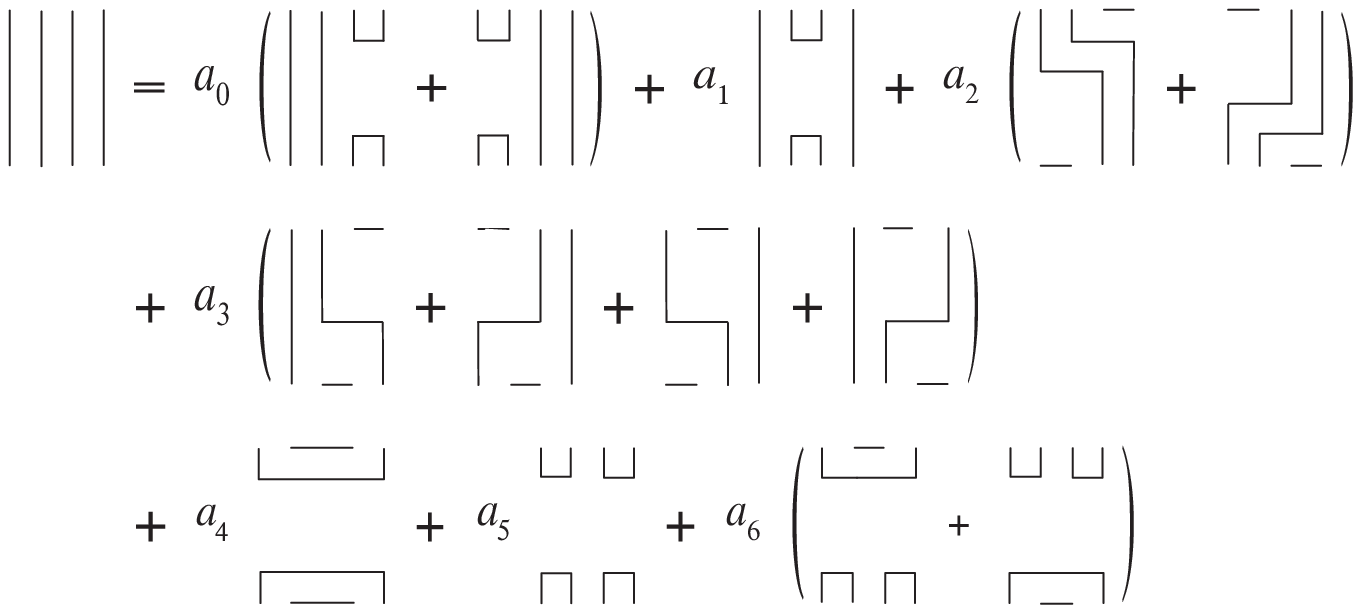}}
\centerline{Figure 6.0} \vskip.2in

In its other incarnation, the 4th Jones-Wenzl's projector relation
\includegraphics[width=.55cm,height=.50cm]{drawinga1.eps} $=0$ looks like
this:

\vskip.2in \epsfxsize=3.5in \centerline{\epsfbox{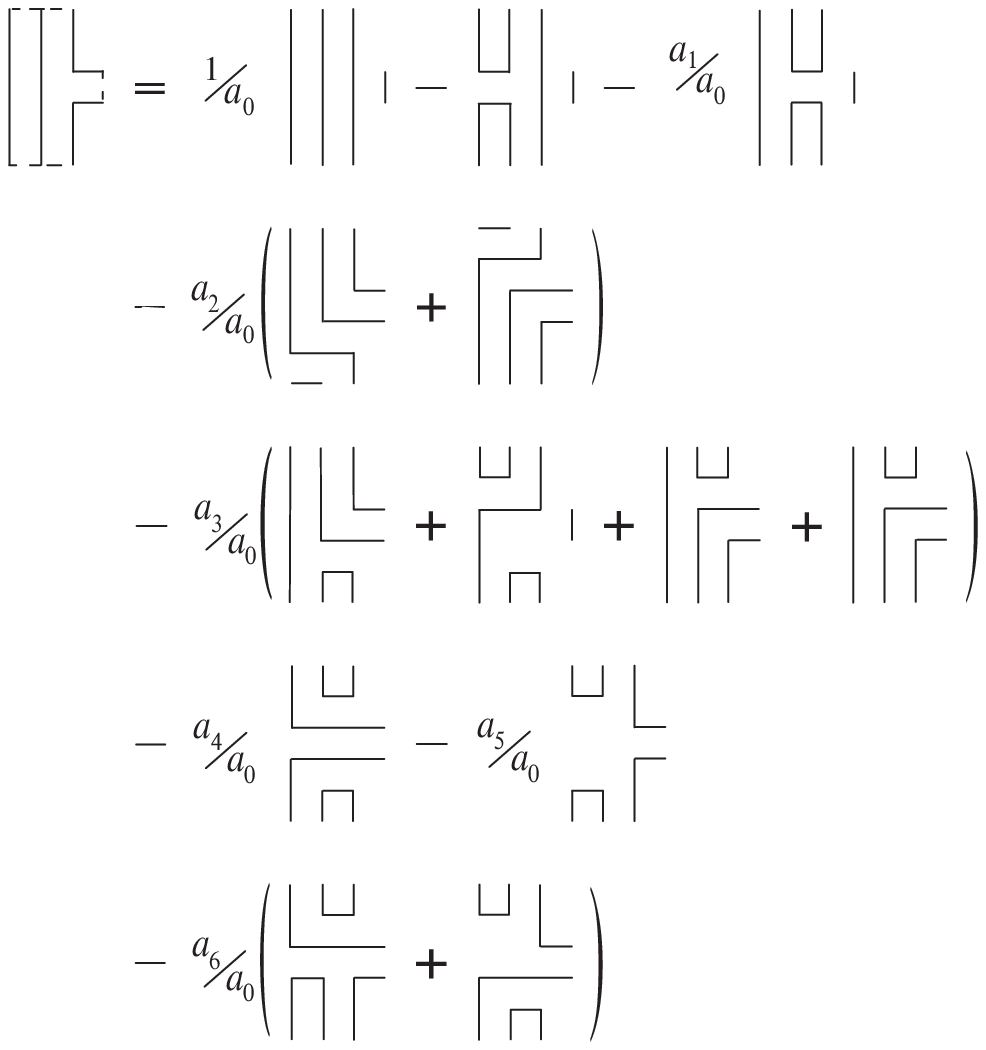}}
\centerline{Figure 6.1} \vskip.2in

The coefficients $a_i$ are rational functors of $d$
which can be computed from the Wenzl's recursion relation for
projectors (see pg. 18 [K, L] or [We]). Figure 6.0 is merely the
lattice counterpart of the more familiar smooth relation, Figure
6.0$^\prime$, which may be applied within any diagram (at $r=5$) whenever
four $1-$ labeled lines are found running parallel. Obviously
Figure 6.1 also has a smooth counterpart.

\vskip.2in \epsfxsize=4in \centerline{\epsfbox{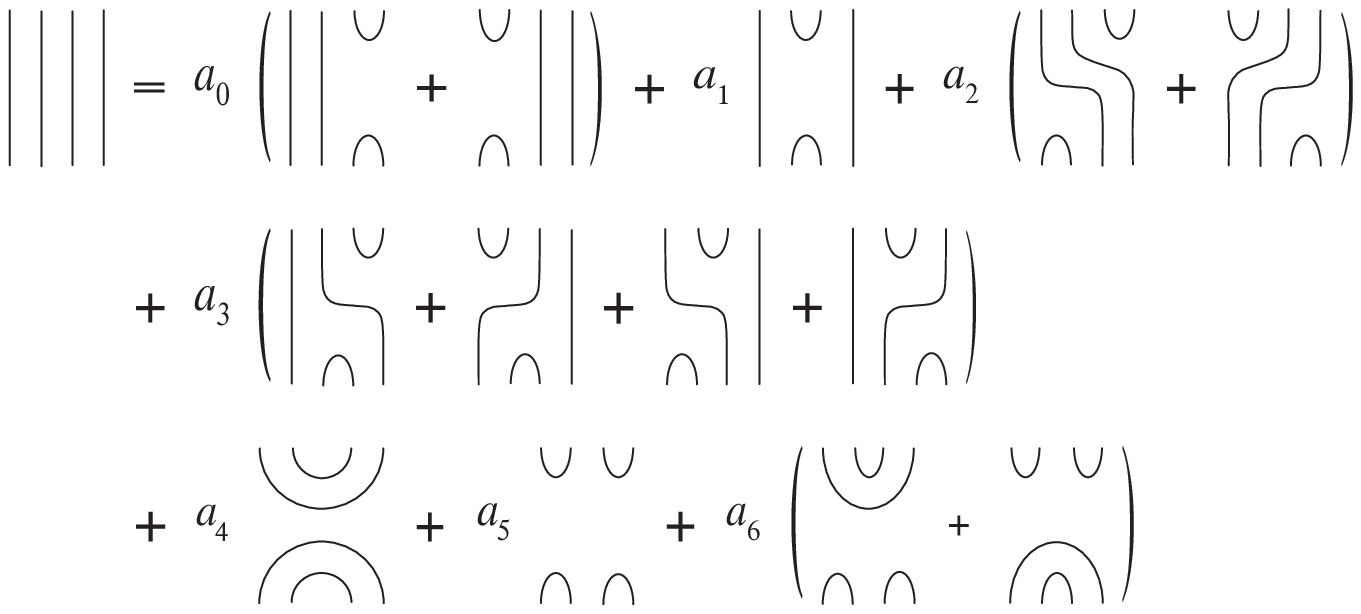}}
\centerline{Figure 6.0$^\prime$} \vskip.2in

The admissibility conditions and the above four classes of
$\lq\lq$equivalences" must be rewritten as operators $A_{i}$ and
$B_{j}$ respectively; collectively denoted $\overline{H}_k$. Let $G$ denote
the ground state of the soon-to-be-defined Hamiltonian
$H=\underset{k}{\S} \ol{H}_k$.  Let $V$ denote the CSr modular
functor of the disk with $3n$ marked points and all labels $=1$.
Via recoupling, we may describe $V$ in the fashion of homology.
Set $P=\C$ [admissible pictures] and write: $V=V_s = P /\sim_s$,
where $\sim_s$ is the smooth-category equivalence relation
corresponding to our four combinatorical equivalences: $\sim_c$.
Lemma 2.1 will prove that under the $\lq\lq$roomy hypothesis"
$\sim_s$ and  $\sim_c$ induce identical equivalence classes of
admissible pictures (which of course are combinatorial objects).
So we may also write $V=V_c = P/ \sim_c$.  Our goal is to tailor
$H$ so that the ground states $g\in G$ correspond bijectively to
linear functionals $\phi:V \la \C$ under the map $\phi\longmapsto
\underset{\underset{\t{admissible pictures}}{p\in}}{\S}\phi
(p)(p)$.  This will identify $G$ with $V^\ast$, but since $V$ has
a canonical nonsingular Hermitian inner product $\big($[Wi] and
[K,L]$\big)$ this also gives an isomorphism $G \cong V$.

The inner product $<p_1, p_2>$ is defined on pictures by
imbedding the disk $D$ into the $(x,y)-$plane, deforming $p_1$
upward rel endpoints and $p_2$ downward rel endpoints.  The
union of the deformed pictures
$\widetilde{p}_1\cup\widetilde{p}_2$ is a (vertically framed) link
in $R^3$ and it Kauffman bracket is  $<p_1, p_2>$. Note that the
vertical framing is singular where $p_1$ and $p_2$ share a common lattice
bonds meeting $\p p_1$ and $\p p_2$; here the convention is to
bend such bonds of $p_2$ slightly clockwise at the endpoints
internal to $D$ and counterclockwise at an endpoint on $\p D$.

The definition of the $A_i$ operators is quite obvious. Consider,
a vertex $v$ in the interior of $R$.  A Hermitian $A_v$ with $4$
indices whose ground state is spanned by classical states of
valence $0$ or $2$ at $v$ is said to enforce
$\lq\lq1-$manifoldness" at $v$.  Clearly the ground state of $A_v$
has dimension $7$.  To enforce, instead, a $\lq\lq$defect" or
marked point labeled by the fundamental representation, $\lq\lq$1"
of $SU(2)$, we would use instead a Hermitian operator
$A_{v}^{\prime}$ with ground state spanned by the four classical
states of valence $=1$ at $v$.

Turning now to $\lq\lq$relations" $B_j$ consider a box $b$ of $R$
centered in a $3 \times 3$ square of boxes:

\vskip.2in \epsfxsize=1.5in \centerline{\epsfbox{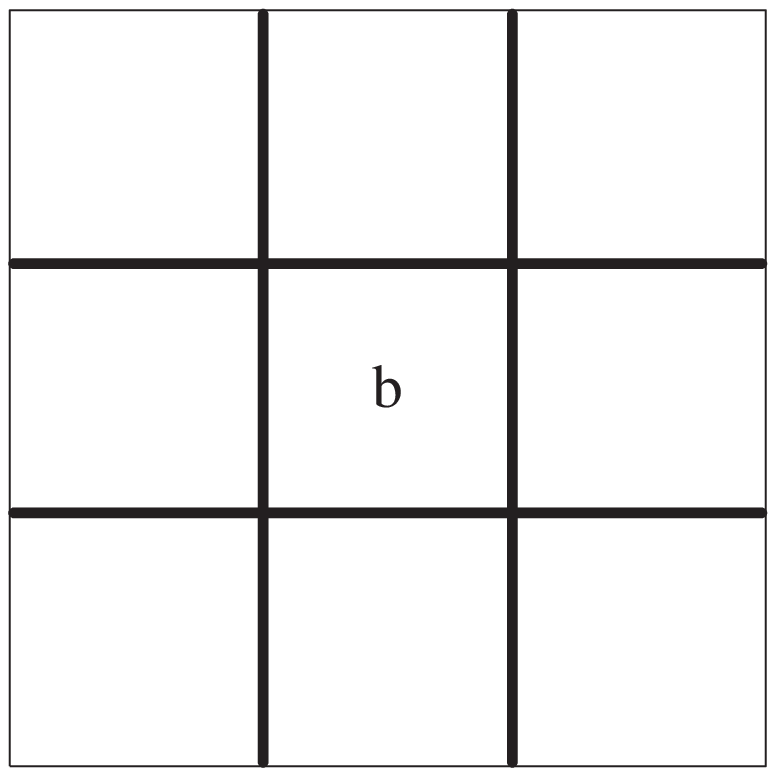}}
\centerline{Figure 7} \vskip.2in

\noindent there are $12$ nonboundary edges $\{e\}$ (shown in
bold). If $\{c\t{'s}\}$ are the nonempty (classical) manifold
configuration of these edges, i.e. valence $\in\{0,2\}$ at each
of the four internal vertices, and iff $c_0$ and $c_{1}=c_{0}
\,\t{xor} \,\p b \,\, \in \{c\t{'s}\}$, set
$d=\f{1}{\sqrt{2}}\left(c_{0} - c_{1}\right)$ and let $\{d\}$ be
the set of such vectors. Let $B_b = \underset{d \e \{d\}}{\S}
\,\, |d><d|$ be the Hermitian operator with $12$ indices on
$(\C^{2})^{\otimes \{e\}}$ whose ground state is orthogonal to
span $\{d\}$. $B_b$ is the operator which $\lq\lq$allows isotopy
across $b$."

To remove circles which bound disks we need, in the presence of
isotopy, only introduce operators which deletes a box.  This
operator may be written as $|\theta ><\theta|$ where $\theta$ is a
unit vector proportional to $|\t{box}>+\big(e^{\pi i/5} + e^{ -\pi
i/5}\big) | \phi >$.

We postpone the definition of the operator corresponding to
figures 3 and 4 since this must involve the dynamics $\lq\lq t$"
of $H_t$. Some trick is needed to avoided adding new levels to our
system to encode $\lq\lq$crossings."

The projector corresponding to \includegraphics[width=.55cm,height=.50cm]{drawinga.eps},
Figure 6.0, requires a $24-$index
operator acting on a $3 \times 3$ grid of edges or $\lq\lq$box" $B$ whose $1-$
dimensional excited state is spanned by the vector obtained by
putting all fourteen term in Figure 6.0 on the left hand side of
the equation.  Similarly the projector corresponding to
\includegraphics[width=.55cm,height=.50cm]{drawinga1.eps}, Figure 6.1 is a 30 index
operator acting on the bonds of a region the shape of l.h.s. in Figure 6.1.  This $\lq\lq$nobby box"
$B^{\prime}$ is a $2 \times 5$ rectangle union an additional small box in the middle of one of the long sides.

Now we turn to the dynamics.  Almost all conditions $H_k$ that
combine to yield $H$ are permanent, only the end point operators
$A_v^\prime$ should change as we execute braiding.  Because of the
technical problem illustrated in figure 4; any lattice resolution
into a superposition of two $1-$manifolds as in figure 3 may cause
collision with other strands.  One way to deal with this problem
is to locate the marked points on a second lattice ${\bf L}^\prime$
consisting of the mid points of the edges in the original Lattice
${\bf L}$ of boxes in $R$. This means that we have to add additional
$2-$index $A$ operators holding equal the two classical states on both halves of
the original edges, i.e. ground state $(A)= ( \mid 00>,| 11>)$, and that
the end point operators $A^\prime_w$ actually occur (with
$2-$dimensional ground states) on the finer lattice ${\bf L}^\prime$,
$w\e {\bf L}^\prime$. The dynamics consists of moving an endpoint
diagonally on ${\bf L}^\prime$, i.e. translating one unit horizontally
or vertically in the structure of ${\bf L}$.  In Figure 8 the endpoint
$w$ is moved horizontally to $w'$ by replacing: $\{A^{\prime}_{w},
A_{w^\prime}\}$ with $\{A_{w},A^{\prime}_{w^\prime}\}$.  If
$w$ and $w'$ are immediately adjacent in ${\bf L}$ the operator swap
will cause the end point to travel around a corner.

\vskip.2in \epsfxsize=4in \centerline{\epsfbox{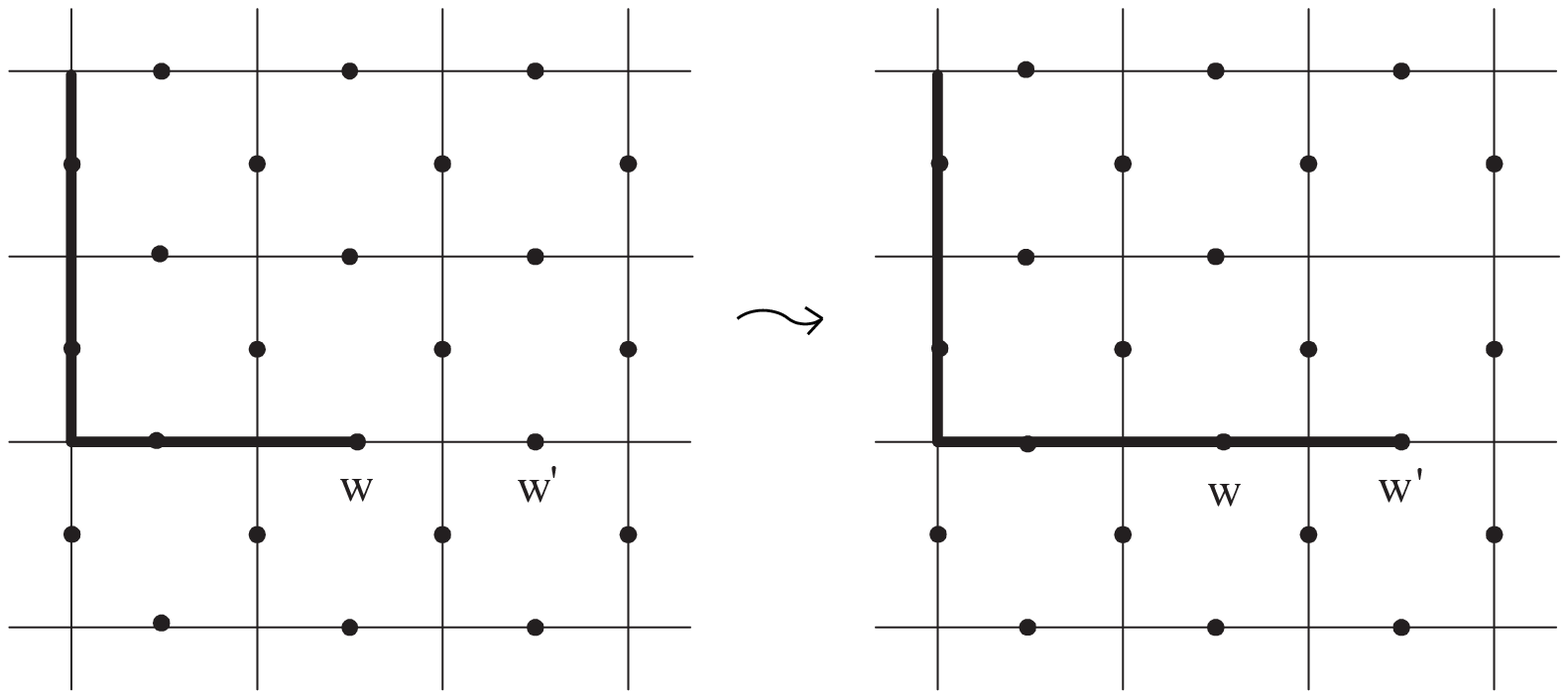}}
\centerline{Figure 8} \vskip.2in

This operator swap can be performed gradually by slowly turning
the appropriate terms on or off.  If the adiabatic theory is
applicable, and following the proof of Lemma 2.1 we discuss the
heuristics for an energy gap (in the theromdynamic limit) for the
family $H_t$, $\psi_t$ will be carried to a unique ground state
$\psi_t$ of $H_{t+1}$.  This ground as a functional on pictures is
identical to $\psi_t$ provided pictures are identified according
to the obvious isotopy rules (and phase rules at endpoints). If the
lattice is refined by a linear factor $L$, tunneling to an
undesired orthogonal ground state $\psi^{\prime}_{t+1}$ should, by
arguments analogous to those for the stability of homology classes
[K2], have amplitudes scaling like $\e (L)=e^{-\Omega (L)}$.  The
mathematical description for adiabatic evolution of the system is
via the natural connection $A$ on the tautological
bundle over the complex Grassmannian $X$: The time evolution of
$G:=\{$ground states $(H_t)\}$ defines a path in $X$ and $A-$transport covers this motion
with a unitary (i.e.isometric) identification $G_0 \equiv G_t$, for all $t\geq0$.
After a braiding $b$ is completed at time $t=T$, the
self-identification $G_0 \equiv G_T$ is the representation of the
braid b.

A ground state $g \in G \subset P$ defines a functional $g^\ast$
on $P$ via orthogonal projection.  The $\{B_j\}$ have
been chosen to correspond to $\sim_c$ precisely so that a unique
extension $\phi$ exists:
\vskip.1in \epsfxsize=.6in \centerline{\epsfbox{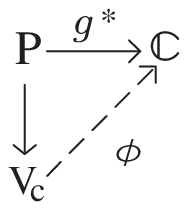}}
\vskip.1in
\noindent and $\phi$ satisfies $g=\S\phi (p)(p)$.  Conversely
given a functional $\phi$ on $V_c$ the $g$ associated to $\phi$
by the formula above lies in the null space of each $B_j$, so in
fact $G=V^\ast_c$.

The important remaining point is to see that after braiding, when
the marked points have been returned to there original sites
set-wise, that the induced transformation on the ground state is
precisely, up to error $\approx \e (L)$, the unitary CS5
representation originally introduced by Jones [J] and studied in
[FLW].  But this follows from the recoupling theory as presented
in [K,L] provided we show that the combinatorial relations  that
we have imposed through Hermitian operators $\{B_j\}$ in fact are
sufficient to span all the relations implied by the infinitely
many smooth relations between pictures, that is $V_s=V_c$.  For
this the following lemma suffices.

\begin{lemma}
Let $\rho =\us{i}{\S} a_i p_i$ be  a linear relation between
admissible combinatorial pictures in $(R, \{3n\})$ which holds
under $\sim_s$, the smooth recoupling theory associated to CS$r$.
Provided that the configuration $\{3n\} \subset R$  is roomy in
the rectangle $R$, the same relation already holds under $\sim_c$.
\end{lemma}

Before proving the lemma let us carry out a simple calculation to
get a feel for how the action of braiding is computed via
recoupling theory.  If the reader wishes to try more complicated
examples, the formulas on pages 93-100 of [K,L] are helpful.  Here
we compute the effect of a braid generator on a vector
$\psi_{\circ}\, \in V:=$CS5 ($3-$punctured disk) where each
boundary component has label$=1$ ( the $2-$dimensional
representation of $sl(2, \C)_q )$ and to account for phase each
boundary has a marked base point.
\[
\psi'_{\circ} \tn{ is the diagram: }
\includegraphics[width=2.5cm,height=.99cm]{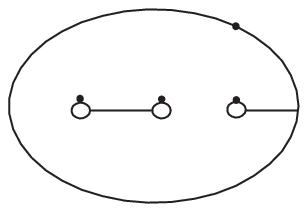},
\]
\noindent which as a labeled tree is:
\[
\psi'_{\circ}=
\includegraphics[width=2.5cm,height=.85cm]{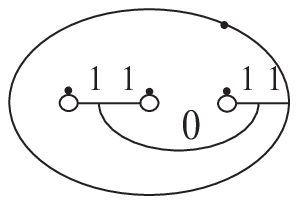}
\]
\noindent Let $b$ be the counterclockwise braided of the right most pair of
punctures. Then $b\psi'_\circ$ is represented by:
\[
\includegraphics[width=2.5cm,height=.85cm]{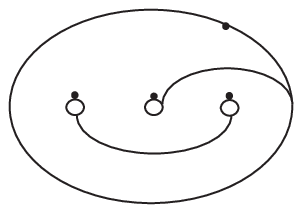}
= \underset{\lq\lq\tn{virtual picture"}}{\includegraphics[width=2.5cm,height=.85cm]{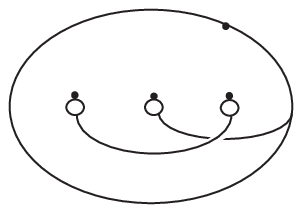}} =
\]
\[
(\ast) \qquad \qquad \quad \quad A \,\,
\includegraphics[width= 2.5cm,height=.85cm]{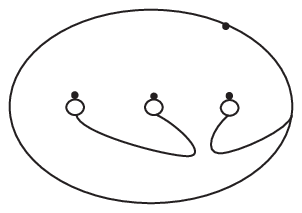} \, \,+ \,\, A^{-1}\,\,
\includegraphics[width= 2.5cm,height=.85cm]{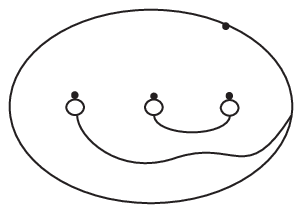}, \tn{ where } A=e^{2 \pi i/10}.
\]

Now $\overset{2}{\includegraphics[width=1.15cm,height=.15cm]{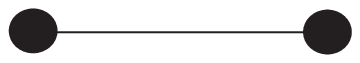}}$ is our notation for the Jones-Wenzl
idempotentent $\overset{2}{\includegraphics[width=1.15cm,height=.20cm]{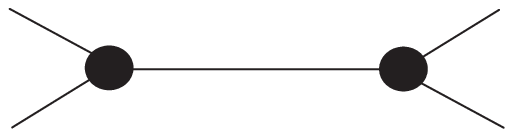}}=
 \f{1}{\sqrt{d^{2}-1}}\left(\includegraphics[width=.75cm,height=.25cm]{drawingf.eps}- \f{1}{d} \supset \subset \right)$
 where $d=-A^2 -A^{-2}$ and, as
usual, the open ends in the above diagrams can be interpreted as
permitting arbitrary (but constant) extension to the outside.
Note: the orthogonality relations
\begin{itemize}
  \item  $< \f{1}{d} \supset \subset, \f{1}{d} \supset \subset > = \f{1}{d^2}
    \, $ \includegraphics[width=.55cm,height=.32cm]{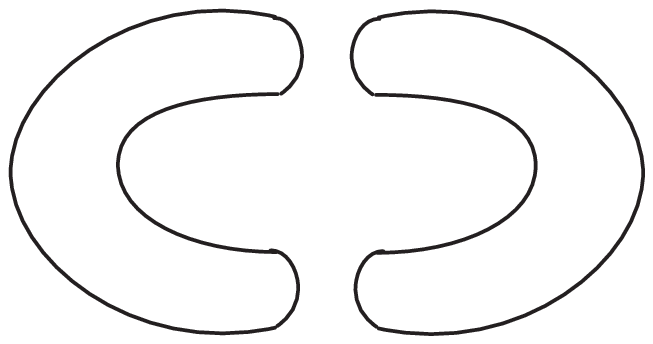} $ = 1 $
  \item  $<\f{1}{\sqrt{d^{2}-1}}\left( \, \includegraphics[width=.75cm,height=.25cm]{drawingf.eps}
    -\f{1}{d} \supset \subset \right), \f{1}{\sqrt{d^{2}-1}}\left(\,
    \includegraphics[width=.75cm,height=.25cm]{drawingf.eps}
    -\f{1}{d}\supset \subset \right) > = \\
    \f{1}{d^{2}-1}\left( \, \includegraphics[width=.55cm,height=.30cm]{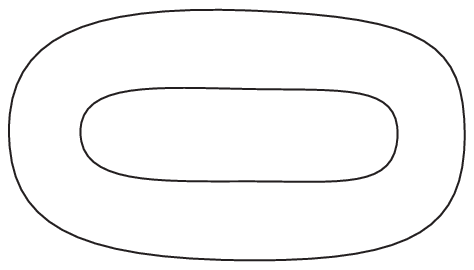}\,
    -\f{2}{d}\, \includegraphics[width=.50cm,height=.33cm]{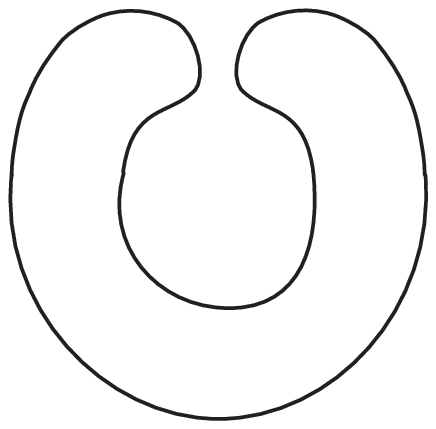}\,
    +\f{1}{d^2}\,\includegraphics[width=.55cm,height=.32cm]{drawingc.eps}
    \right)= \f{1}{d^2 -1}\left(d^2 - \f{2}{d} d +\f{1}{d^2} d^2\right)=1$, and
  \item $<\f{1}{\sqrt{d^{2}-1}}\left( \, \includegraphics[width=.75cm,height=.25cm]{drawingf.eps}
    -\f{1}{d} \supset \subset \right),\f{1}{d} \supset\ \subset > =
    \f{1}{d\sqrt{d^{2}-1}}\left( \, \includegraphics[width=.50cm,height=.33cm]{drawingd.eps}
    -\f{1}{d} \, \includegraphics[width=.55cm,height=.32cm]{drawingc.eps} \right) = \\
    \f{1}{d\sqrt{d^{2}-1}}\left( d-\f{1}{d}d^2\right) = 0$
\end{itemize}
Normalizing, $\psi_\circ =\f{1}{d} \psi'_t$ is a unit vector,
$<\psi_\circ , \psi_\circ > = 1$.

From the definition of $\overset{2}{\includegraphics[width=1.15cm,height=.15cm]{drawingg.eps}}$ we have:
\includegraphics[width=.75cm,height=.25cm]{drawingf.eps}$=\sqrt{d^{2}-1}$
$\overset{2}{\includegraphics[width=1.15cm,height=.15cm]{drawingg.eps}}- \f{1}{d}\supset\subset$.
So we use this to expand the two parallel lines in the second term
of $(\ast)$ to get:
\begin{align*}
b\psi_\circ =& A \psi_\circ + \f{A^{-1}}{d} \left( \includegraphics[width=2.5cm,height=.85cm]{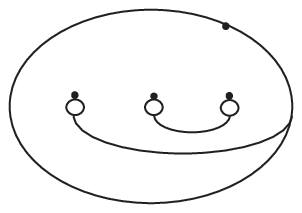}\right)\\
=& A \psi_\circ + \f{A^{-1}}{d} \left( \sqrt{d^{2} -1} \quad \includegraphics[width=2.5cm,height=.85cm]{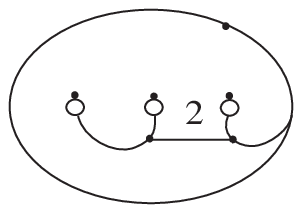}
+ \, \f{1}{d}\quad \includegraphics[width=2.5cm,height=.85cm]{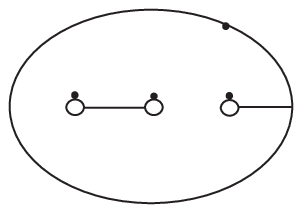} \right) (**)\\
=& A \psi_\circ + \f{\sqrt{d^{2}-1}}{d}
\,A^{-1} \psi_2 +\f{A^{-1}}{d} \psi_\circ\\
\tn{where } \psi_2  := & \,
\includegraphics[width=2.5cm,height=.85cm]{drawinga12.eps}, \\
 =& \left(A +\f{A^{-1}}{d}\right)\psi_\circ + \f{\sqrt{d^{2}-1}}{d} \,A^{-1} \psi_2
\end{align*}
As a check on unitarity note that under the sequelinear pairing,
\begin{align*}
<b \psi_\circ, b\psi_\circ > = &\left(A +\f{A^{-1}}{d}\right)\left(A^{-1} +\f{A}{d}\right) +
\left(\f{\sqrt{d^{2}-1}}{d} \,A^{-1} \f{\sqrt{d^{2}-1}}{d}A\right) \\
 = & 1 + \f{A^{2} + A^{-2}}{d} + \f{1}{d^{2}} + \f{d^{2}-1}{d^2} =1
\end{align*}

\noindent {\bf{Proof of Lemma 2.1.}}  The argument is based on the
Birkhoff curve shortening principle where by a family of imbedded
arcs and circles can be $\lq\lq$pulled tight" to a shorter
geodesic position without crossings developing.  We work
combinatorically. By the $\lq\lq$roomy hypothesis" there is an
$r-$collared tree $T:=T_r \subset R$.  Assign a positive weight
$w(\beta)$ to each bond $\beta$ (or $1-$cell) of the cellutation of $R$
so that $w$ grows rapidly with distance to $T$:  as a good first approximation, we may take
$w(\beta)=10^{\# (\beta)}$ where $\#(\beta)=$minimum number of
bonds joining $\beta$ to $T$.  Now for any (classical) picture
$p_i$ define its length $L(p_i)= \underset{\beta \e p_i}{\S}
w(\beta)$. Permitting combinatorial isotopy (rel the marked
points) and the removal of small circles, but not the
undercrossing,
\includegraphics[width=.55cm,height=.50cm]{drawinga.eps}, or
\includegraphics[width=.55cm,height=.50cm]{drawinga1.eps}
relations, we may pull $p_i$ tight by local moves to equivalent
pictures (up to a scalar) which steadily reduce $L(p_i)$ until a
local minimum is reached.  Call this step $\lq\lq$pull tight".
Because of our weight function $w$, the new $p_i$ will try to lie
mainly in a small neighborhood of $T$, and only occupancy of the
bonds close to $T$ will force parts of the picture to lie farther
away.  Also the picture, seeking to occupy the bonds near $T$
efficiently will have its strands running parallel to $T$ in
$(r-2)\times(r-2)$ lattice blocks $\beta$ a thwart the middle
$r-$bonds of each of the distinguished length $3r$ segments of
$T$.  Also at the trivalent vertices of $T$ near which
sufficiently many strands pass, we would like to see copies of
l.h.s. Figure 6.1.  This will be true up to a small isotopy
(across a few boxes) and can be made true on the nose by
modifying the weight function $w$ by adding a small term
proportional to the distance from each trivalent vertex out to a
distance $r$ from that vertex. Now apply
\includegraphics[width=.55cm,height=.50cm]{drawinga.eps} at some
site in a $B$ or
\includegraphics[width=.55cm,height=.50cm]{drawinga1.eps} at some
site $B^{\prime}$ if the opportunity presents.  This breaks $p_i$
into $\underset{j}{\S}\,  b_{ij} q_{ij}$ and for all
$j$,  $L(q_{ij})<L(p_i )$. Pull tight again to remove the slack
created by the $\lq\lq$turn arounds" in Figure 6.0  or 6.1.
Alternate pulling tight with applications of
\includegraphics[width=.55cm,height=.50cm]{drawinga.eps} or \includegraphics[width=.55cm,height=.50cm]{drawinga1.eps}
until no further reductions in length can be made in this way.
Call this cycle $\lq\lq$pull and cut".  With a slight abuse of
notation let $q_{ij}$ denote one of the terminal classical states
of this process.  Now allow a single $\lq\lq$under crossing" move
(Figure 3) to further reduce $L(q_{ij})$ if such a move is
available.  Now alternated the $\lq\lq$pull and cut" cycle with
single under crossing moves until no daughter picture (still
denote $q_{ij}$) can have its length reduced by further iteration
of this process.

Manifestly, all the daughter pictures $q_{ij}$ now lie in $N(T)$
and pass through all boxes $B$ parallel to $T$ and with $3$
or few strands and pass through each $B^{\prime}$ in a standard way
according to some admissible triple as explained below.  Note
that $(3, 3, 2)$ is not admissible.  At this point it is simple to
formally reorganize the term of this sum  $\underset{ij}{\S}\,
{b}_{ij}, q_{ij}$ as $\S c_\ell T_\ell$ where
$T_\ell$ is an admissible labeling of $T$.  As explained in
[K,L], the leaves and root $T$ of $t$ are always labeled  by $1$
(this is our choice) and the admissibility condition says that
other edges (i.e. components of the intrinsic $1-$skeleton of
$T$) are labeled by $a, b, c, d, \ldots$ taken from $\{0, 1, 2,
\ldots, r-2\}$ so that at each trivalent vertex of $T$ the
following relations hold on the triple of incident labels $a, b,$
and $c$:
\begin{align*}
  a & \leq b + c \\
  b & \leq c + a \\
  c & \leq a + b \\
  a + b + c &\equiv 0 \,\tn{(mod 2), and }\\
  a + b + c &< 2(r-1)=8.
\end{align*}
An admissible labeling $T_\ell$ is interpreted as a linear
combination of pictures by replacing each edge with the
Jones-Wenzl projector corresponding to its label.  The set of
admissible labeled trees $\{T_\ell \}$ is an orthogonal basis for
the modular functor $V_s (R, \{3n\})$, defined topologically
using the smooth equivalence relation.  (The subscript $s$ is to
emphasis that the smooth relations are used in this definition;
of course $V_s = V$.)

Because of the assumed  $p = \underset{i,j}{\S}\, \, b_{ij},
q_{ij}= 0\, \in V_s (R, \{3n\})$, $c_\ell =0$
for all admissible $\ell$.  But each $q_{ij}$ is an imbedded arc
pairing $x$ of the $\{$leaves $\cup$ root$\}$ in $N(T)$
satisfying the additional admissibility restriction at each
trivalent vertex of $T$.  Such pairings are an alternative
(though not orthogonal) basis for the modular functor $V(R,
\{3n\})$ so collected in this basis we have for each pairing type
$x$, we have $\S\, \, b^{x}_{ij}, q^{x}_{ij}= 0$ where
  \begin{align*}
    b^{x}_{ij} = b_{ij} &\text{ if } q_{ij} \text{ has type } = x \\
     = 0 &\text{ if } q_{ij} \text{ has type } \neq x.
  \end{align*}
  But all $q_{ij}$ of a fixed type are clearly combinatorally
  equivalent $(\sim_c)$.  Thus we have found a combinatorial path
  through applications of $(\sim_c)$ from $p$ to the empty
  picture, or more precisely to a sum of zero times various
  pictures.  $\square$

Unlike [K2] the individual summands of $H$ do not commute. The
ground states of $H$ has been computed topologically, however the spectrum
spec$(H)$ is less accessible. The most important question
is the existence of an energy gap above the ground state which is
constant under lattice refinement, $L \la \infty$, i.e. in the
thermodynamic limit. The following heuristics motivate the
conjectured energy gap.

In finite classical systems such as random walk on a graph
diffusion time is well known to scale inversely with the spectral
gap of the Laplacian.  Similarly, in some simple quantum
mechanical systems where exact calculation is possible, the energy
gap scales inversely to the diffusion time between classical
states.  In [K2] where direct calculation yields an energy gap
above the ground state, the classical states are cycles and the
$\lq\lq$diffusion" is through elementary bordisms.  Since we have
set up our ground state to be analogous to homology: $G \cong P/
\sim_c$ with pictures playing the role of cycles and our $\{B_j\}$
playing the role of bordisms we expect similar diffusion
properties and hence an energy gap.  In lemma 2.1 the proof shows
that equivalent pictures $p_1$ and $p_2$ are connected by a
$\lq\lq$path" $\gamma$ of deformations ($\lq\lq$down" from $p_1$
to a neighborhood of $T_r$ and then back up to $p_2$).  Rapid
diffusion corresponds to observing that there are a plethora of
such paths and in fact the procedure for finding $\gamma$ is
highly under determined.  More difficult would be a rigorous
implication between diffusion and spec$(H)$.  Extending the
analogy with [K2], in both cases when the lattice is refined by a
factor of $L$, a sequence of $O(L)$ local operators is required
to transform between a pair of orthogonal ground states. So given
the existence of an energy gap, the Hamiltonian $H$ will be stable
to order ${O}(L)$ in perturbation theory; formally corresponds
to tunneling amplitudes between orthogonal ground states which
scale like $e^{-\Omega(L)}$.

There are several important open questions. The first is a
rigorous treatment of the energy gap, but this is probably too
difficult in the present model. Another is how to deal with errors
in the form of actual rather than $\lq\lq$virtual" excitation
which have already been discussed in the context of tunneling. Can
a coupling to a could bath repair such errors or are more active
measures required?                                                                                                                                  For example, can broken endpoint pairs of a
$1-$manifold find each other and cancel through some imposed
attraction (as suggested by Dan Gottesman in conversations) or
merely through random walk? Nearby error pairs may be more serious
in CS5 than in the toric codes since isotopy class not just
homology needs to be preserved; the wrong reconnection pairing
would result in an unrecoverable error. To make this unlikely,
should additional terms be included into our Hamiltonian $H$ which
could force distinct strands to be widely separated?  This would
put more weight on the simpler pictures, which are the ones that
the quantum medium can most easily correct if damaged.

Kitaev's very general notion of quantum media with its several
antecedents in the study of quantum statistical mechanics looks
likely to become a central object of study shared between
theoretical physics, solid state physics, and topology.  The main
disappointment of the present investigation  is the complexity of
the local Hamiltonian $H$ used to construct stable universal
\underline{topological} quantum computation.  One sees no easy
road to radically simplifying it and still obtaining an exact
description of CS5.  However another path may be open. In our
discussions, Kitaev has suggested (also see page 46 [P]) that
simpler lattice Hamiltonians may renormalize in the scaling limit
to topological modular functors.  Perhaps the most interesting
topological theories, such as CS5, because of their simplicity
will have large $\lq\lq$basins of attraction" under
renormalization and that identifiable universality classes of
quantum media may not only exist mathematically but may even lie
within the reach of engineers.

\end{document}